\documentclass[twocolumn,english,superscriptaddress,citeautoscript,showpacs,preprintnumbers,amsmath,amssymb,prl,floatfix,footinbib]{revtex4}
\usepackage[scaled=2]{helvet}
\usepackage[T1]{fontenc}
\usepackage[latin9]{inputenc}
\usepackage{amsmath}
\usepackage{amssymb}
\usepackage{graphicx}

\makeatletter

\providecommand{\tabularnewline}{\\}

\@ifundefined{textcolor}{}
{%
 \definecolor{BLACK}{gray}{0}
 \definecolor{WHITE}{gray}{1}
 \definecolor{RED}{rgb}{1,0,0}
 \definecolor{GREEN}{rgb}{0,1,0}
 \definecolor{BLUE}{rgb}{0,0,1}
 \definecolor{CYAN}{cmyk}{1,0,0,0}
 \definecolor{MAGENTA}{cmyk}{0,1,0,0}
 \definecolor{YELLOW}{cmyk}{0,0,1,0}
}

\usepackage[scaled=2]{helvet}\usepackage{amstext}

\makeatletter
\@ifundefined{textcolor}{}{\definecolor{BLACK}{gray}{0}\definecolor{WHITE}{gray}{1}\definecolor{RED}{rgb}{1,0,0}\definecolor{GREEN}{rgb}{0,1,0}\definecolor{BLUE}{rgb}{0,0,1}\definecolor{CYAN}{cmyk}{1,0,0,0}\definecolor{MAGENTA}{cmyk}{0,1,0,0}\definecolor{YELLOW}{cmyk}{0,0,1,0}}

\usepackage{dcolumn}
\usepackage{bm}
\usepackage{times}
\usepackage{hyperref}
\hypersetup{colorlinks=true,linkcolor=red,citecolor=blue}
\makeatother

\usepackage{babel}

\makeatother

\usepackage{babel}
\begin{document}

\title{Coexistence of Superconductivity and Ferromagnetism in P-doped EuFe$_{2}$As$_{2}$}

\author{S. Nandi}

\email{s.nandi@fz-juelich.de}

\affiliation{Jülich Centre for Neutron Science JCNS and Peter Grünberg Institut
PGI, JARA-FIT, Forschungszentrum Jülich GmbH, D-52425 Jülich, Germany}

\affiliation{Jülich Centre for Neutron Science JCNS, Forschungszentrum Jülich
GmbH, Outstation at MLZ, Lichtenbergstraße 1, D-85747 Garching, Germany}

\author{W. T. Jin}

\affiliation{Jülich Centre for Neutron Science JCNS and Peter Grünberg Institut
PGI, JARA-FIT, Forschungszentrum Jülich GmbH, D-52425 Jülich, Germany}

\affiliation{Jülich Centre for Neutron Science JCNS, Forschungszentrum Jülich
GmbH, Outstation at MLZ, Lichtenbergstraße 1, D-85747 Garching, Germany}

\author{Y. Xiao}

\affiliation{Jülich Centre for Neutron Science JCNS and Peter Grünberg Institut
PGI, JARA-FIT, Forschungszentrum Jülich GmbH, D-52425 Jülich, Germany}

\author{Y. Su}

\affiliation{Jülich Centre for Neutron Science JCNS, Forschungszentrum Jülich
GmbH, Outstation at MLZ, Lichtenbergstraße 1, D-85747 Garching, Germany}

\author{S. Price}

\affiliation{Jülich Centre for Neutron Science JCNS and Peter Grünberg Institut
PGI, JARA-FIT, Forschungszentrum Jülich GmbH, D-52425 Jülich, Germany}

\author{D. K. Shukla}

\affiliation{Deutsches Elektronen-Synchrotron DESY, D-22607 Hamburg, Germany}

\affiliation{UGC DAE Consortium for Scientific Research, Khandwa Road, Indore
01, India}

\author{J. Strempfer}

\affiliation{Deutsches Elektronen-Synchrotron DESY, D-22607 Hamburg, Germany}

\author{H. S. Jeevan}

\affiliation{I. Physikalisches Institut, Georg-August-Universität Göttingen, D-37077
Göttingen, Germany}

\author{P. Gegenwart}

\affiliation{I. Physikalisches Institut, Georg-August-Universität Göttingen, D-37077
Göttingen, Germany}

\author{Th. Brückel}

\affiliation{Jülich Centre for Neutron Science JCNS and Peter Grünberg Institut
PGI, JARA-FIT, Forschungszentrum Jülich GmbH, D-52425 Jülich, Germany}

\affiliation{Jülich Centre for Neutron Science JCNS, Forschungszentrum Jülich
GmbH, Outstation at MLZ, Lichtenbergstraße 1, D-85747 Garching, Germany}
\begin{abstract}
The magnetic structure of the Eu$^{2+}$ moments in the superconducting
EuFe$_{2}$(As$_{1-x}$P$_{x}$)$_{2}$ sample with \emph{x}\,=\,0.15
has been determined using element specific x-ray resonant magnetic
scattering. Combining magnetic, thermodynamic and scattering measurements,
we conclude that the long range ferromagnetic order of the Eu$^{2+}$
moments aligned primarily along the \textbf{c} axis coexists with
the bulk superconductivity at zero field. At an applied magnetic field
$\geq0.6$\,T, superconductivity still coexists with the ferromagnetic
Eu$^{2+}$ moments which are polarized along the field direction.
We propose a spontaneous vortex state for the coexistence of superconductivity
and ferromagnetism in EuFe$_{2}$(As$_{0.85}$P$_{0.15}$)$_{2}$.
\end{abstract}
\maketitle

\section{Introduction}
The discovery of the iron-based superconductors \cite{kamihara_08}
a few years ago has stimulated tremendous research interests worldwide
in unconventional high-\emph{T}$_{\textup{C}}$ superconductivity
\cite{Johnston}. Most of the research on the Fe-based superconductors
has focused on mainly four systems, (1) the quaternary \textquotedblleft{}1111\textquotedblright{}
systems, \emph{R}FeAsO$_{1-x}$F$_{x}$ (\emph{R}\,=\,La, Nd, Sm,
or Pr, etc.) with \emph{T}$_{\textup{C}}$ as high as 56 K \cite{kamihara_08,Takahashi_08,chen_08,Ren_08},
(2) the ternary \textquotedblleft{}122\textquotedblright{} systems,
\emph{A}Fe$_{2}$As$_{2}$ (\emph{A}\,=\,Ba, Ca, Sr, or Eu etc) with
\emph{T}$_{\textup{C}}$ upto 38 K, \cite{rotter_08,Jeevan_08,Sasmal_08},
(3) the binary \textquotedblleft{}11\textquotedblright{} system (e.\,g.
FeSe) \cite{Hsu_08} with \emph{T}$_{\textup{C}}$\,\textasciitilde{}\,18
K and (4) the ternary ``245'' systems, \emph{A}$_{2}$Fe$_{4}$Se$_{5}$
(\emph{A} = K, Rb, Cs) with \emph{T}$_{\textup{C}}$\,\textasciitilde{}\,30
K \cite{Guo_10}. Superconductivity can be achieved in all the above
compounds in different ways, for example, either by electron or hole
doping in the Fe-As layers \cite{Leithe_08,Athena_08} or by isovalent
substitution \cite{Schnelle_09,wang_09,Jiang_09}. Internal chemical
pressure by isovalent substitution of arsenic with phosphorus \cite{wang_09,Jiang_09}
or external hydrostatic pressure can also give rise to superconductivity
\cite{Miclea_09,Tokiwa_12}.

EuFe$_{2}$As$_{2}$ is an interesting member of the \textquotedblleft{}122\textquotedblright{}
family since the \emph{A} site is occupied by Eu$^{2+}$, which is
an \emph{S}-state rare-earth ion possessing a 4\emph{f}$^{7}$ electronic
configuration with the electron spin \emph{S} = 7/2 \cite{Marchand_1978}.
EuFe$_{2}$As$_{2}$ exhibits a spin density wave (SDW) transition
in the Fe sublattice concomitant with a structural phase transition
at 190 K. In addition, Eu$^{2+}$ moments order in an A-type antiferromagnetic
(AFM) structure at 19 K (ferromagnetic layers ordered antiferromagnetically
along the\emph{ }\textbf{c} axis) \cite{Martin_09,Xiao_09,Xiao_10}.
Superconductivity can be achieved in this system by substituting Eu
with K or Na (Refs.\,{[}\onlinecite{Qi_2008, Jeevan_08}{]}), As
with P (Ref. \onlinecite{Ren_09}), and upon application of external
pressure (Refs.\,{[}\onlinecite{Miclea_09, Terashima_09,Tokiwa_12}{]}).

Superconductivity and magnetism are two antagonistic phenomena since
the superconducting state expels external magnetic flux. Nevertheless,
superconductivity in the pnictides and cuprates is always found in
close proximity to an antiferromagnetic order and the superconducting
pairing is believed to be mediated by the antiferromagntic spin fluctuations
\cite{Johnston}. Most surprising is the coexistence of ferromagnetism
and superconductivity as recently proposed by many groups for the
P-doped EuFe$_{2}$As$_{2}$ samples \cite{cao_11,Ahmed_10,Nowik,Zapf_11,Zapf_13}.
Based on Mössbauer studies on superconducting polycrystalline samples,
Nowik \emph{et al}. \cite{Nowik} concluded that the Eu$^{2+}$ moments
are aligned ferromagnetically along the \textbf{c} axis with a possible
tilting angle of 20$^{\circ}$ from the \textbf{c} axis. Zapf \emph{et
al}. also \cite{Zapf_11} concluded based on macroscopic measurements
that the Eu$^{2+}$moments in EuFe$_{2}$(As$_{1-x}$P$_{x}$)$_{2}$
order in a canted A-type antiferromagnetic structure with the spin
component along the \textbf{c} direction being ferromagnetically aligned.
The small in plane component of the Eu$^{2+}$moments in the A-type
AFM structure undergoes a spin glass transition where the moments
between the layers are decoupled \cite{Zapf_13}.

For a magnetic superconductor with rare-earth moments, several theoretical
studies claim that the superconductivity can coexist with several
forms of the magnetic states, namely, (a) \textquotedblleft{}cryptoferromagnetism\textquotedblright{}
(which is a ferromagnetic state with small domains, smaller than the
superconducting coherence length) \cite{Anderson_59} or (b) transverse
amplitude modulated collinear antiferromagnetic structure or (c) spiral
antiferromagnetic structure or (d) with a spontaneous vortex state
of the magnetic moments. A spontaneous vortex state or a self-induced
vortex state is a new state of matter in which the two competing orders,
superconductivity and ferromagnetism, coexist due to the lower free
energy of the combined states compared to the individual ones \cite{Greenside_81}.
The Pure ferromagnetic state is least preferred. These results clearly
show the importance of the alignment for the rare-earth moments in
the superconducting samples.

To the best of our knowledge, for the superconducting EuFe$_{2}$(As$_{1-x}$P$_{x}$)$_{2}$
single crystal samples, direct microscopic evidence for the proposed
ferromagnetic and/or antiferromagnetic structure is still lacking.
Due to the strong neutron absorption of Eu together with the small
sample mass of the P-doped single crystals, the magnetic structure
determination in EuFe$_{2}$(As$_{1-x}$P$_{x}$)$_{2}$ via neutron
diffraction is considerably more challenging than that of other members
of the new superconductors. The only attempt was made on a powder
sample of the non-superconducting EuFe$_{2}$P$_{2}$ where it was
concluded that the Eu$^{2+}$ moments order ferromagnetically with
a canting angle of 17$^{\circ}$ from the \textbf{c} axis \cite{Ryan_11}.
Here we report on the first element-specific x-ray resonant magnetic
scattering (XRMS) studies of the superconducting EuFe$_{2}$(As$_{1-x}$P$_{x}$)$_{2}$
to explore the details of the magnetic structure of the Eu$^{2+}$
moments. Our resonant scattering experiments show that the Eu$^{2+}$
moments order ferromagnetically along the \textbf{c} axis at zero
field and undergo a transition into a field induced ferromagnetic
state along the applied magnetic field direction for applied magnetic
fields $\geq$ 0.6\,T. Both the zero and applied magnetic field ferromagnetic
order of the Eu$^{2+}$ moments coexist with the bulk superconductivity.
\section{Experimental Details}
Single crystals of EuFe$_{2}$(As$_{1-x}$P$_{x}$)$_{2}$ with $x=0.05$
and $x=0.15$ were grown using FeAs flux \cite{Jeevan_11}. For the
scattering measurements and for the superconducting composition $x=0.15$,
an as-grown right isosceles triangular shaped single crystal with
a base of approximately 2 mm and a thickness of 0.1 mm was selected.
The \emph{same} crystal was used for all the macroscopic characterizations
presented in this communication. For the non-superconducting $x=0.05$
sample, a crystal of approximate dimensions of 2$\times$2$\times0.1$\,mm$^{3}$
was chosen. The surface of both single crystals were perpendicular
to the $c$ axis. The XRMS experiments were performed at the Eu L$_{3}$-edge
at beamline P09 at the PETRA III synchrotron at DESY \cite{Strempfer_13}.
The incident radiation was linearly polarized parallel ($\pi$-polarization)
and perpendicular ($\sigma$-polarization) to the horizontal and vertical
scattering planes for the 15\% and 5\% doped samples, respectively.
The spatial cross section of the beam was 0.2\,(horizontal)$\times$0.05\,(vertical)
mm$^{2}$. Copper Cu\,(2\,2\,0) was used at the Eu L$_{3}$ absorption
edge as a polarization and energy analyzer to suppress the charge
and fluorescence background relative to the magnetic scattering signal.
The sample was mounted at the end of the cold finger of a cryomagnet
with {[}2 1 0{]}$_{T}$-{[}0 0 1{]}$_{T}$ plane coincident with the
scattering plane for the 15\% doped sample. The magnetic field was
applied along the {[}1 $\bar{2}$ 0{]} direction which is perpendicular
to the scattering plane. The 5\% doped sample was measured inside
a closed cycle Displex cryogenic refrigerator with {[}1 1 0{]}$_{T}$-{[}0
0 1{]}$_{T}$ as the scattering plane. Measurements at P09 were performed
at temperatures between 5 and 180\,K. For convenience, we will use
tetragonal (\emph{T}) notation unless otherwise specified.

\begin{figure}
\centering{}\includegraphics[width=0.42\textwidth]{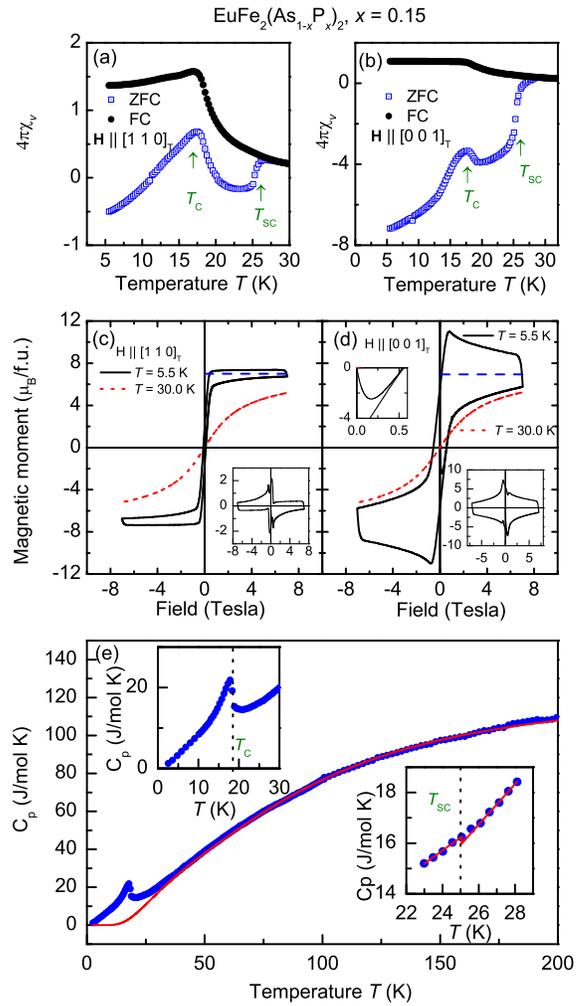}\\
 \caption{\label{fig1}(a) and (b) Temperature dependencies of the magnetic
susceptibility measured on heating of the zero-field cooled (ZFC)
and field cooled (FC) sample at an applied magnetic field of 1\,mT
along the crystallographic {[}1 1 0{]}$_{T}$ and {[}0 0 1{]}$_{T}$
directions, respectively. (c) and (d) \emph{M}-\emph{H} curves for
magnetic fields parallel and perpendicular to the \textbf{c} axis
at \emph{T} = 5 K (below magnetic and superconducting transitions)
and 30 K (above superconducting and magnetic transitions). Horizontal
dashed lines in both figures denote fully saturated moment of Eu$^{2+}$.
Lower insets for both figures show the hysteresis curves after subtraction
of the ferromagnetic contribution as described in the text. The upper
inset of the Fig. 1 (d) shows details of the \emph{M-H} dependence
in the low filed region. (e) Temperature dependence of the specific
heat. Upper and lower insets show details near the magnetic ordering
of the Eu$^{2+}$ and the superconducting transition, respectively.
The solid curve represents the fit using Debye and Einstein contributions
for the lattice part of the specific heat. The lattice part was subtracted
from the total heat capacity to calculate the entropy release at \emph{T}$_{\textup{C}}$.}
\end{figure}

\section{Experimental Results}

\subsection{Macroscopic Characterizations}
\begin{figure}
\begin{centering}
\includegraphics[width=0.42\textwidth]{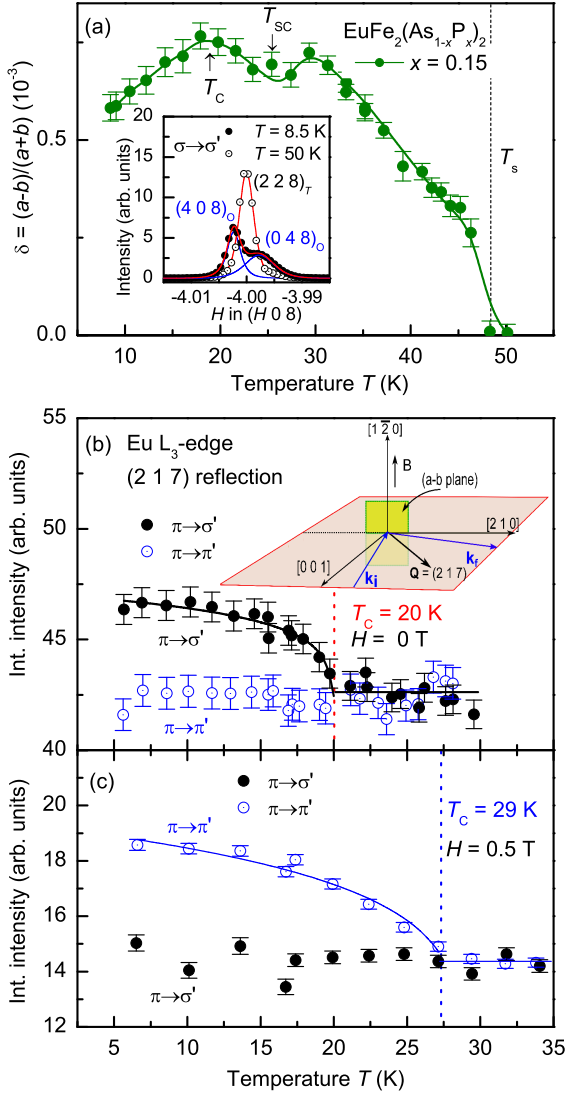}
\par\end{centering}

\centering{}

\begin{centering}
\caption{\label{fig2}(a) Temperature dependence of the orthorhombic distortion
for the \emph{x}\,=\,0.15 sample. The inset shows ($\xi$~$\xi$~0)$_{T}$
scans through the (2 2 8)$_{T}$ position above and below the structural
phase transition. The lines represent fits to the data using either
one (red) or two (blue) Lorentzian squared peaks. (b) Temperature
dependence of the (2 1 7) reflection in both the $\pi\rightarrow\sigma'$
and $\pi\rightarrow\pi'$ scattering geometries at zero filed. The
schematic shows the used scattering geometry. (c) Same as (b) but
in an applied magnetic field of 0.5 T. The temperature dependencies
were measured at the peak energy ($\sim$ 6.973 keV) of the resonance
enhancement observed in the energy scans. }

\par\end{centering}

\centering{}\label{figure2}
\end{figure}

Figure\,\ref{fig1} (a-b) and (c-d) show magnetic susceptibility
(\emph{M-T}) and isothermal magnetization (\emph{M-H}) of the $x=0.15$
sample, respectively, measured for magnetic fields parallel and perpendicular
to the \emph{c} axis using a Quantum Design (SQUID) magnetometer.
Zero field cooled magnetization becomes negative for both field directions
at \emph{T}$_{\textup{\textup{SC}}}$ = 25 K, signifying a superconducting
transition at this temperature. Upon cooling towards the onset of
Eu$^{2+}$ ordering at \emph{T}$_{\textup{C}}$ = 19 K, the superconducting
signal is first weakened, before it becomes more pronounced at temperatures
below \emph{T}$_{\textup{C}}$. Superconductivity wins over the Eu$^{2+}$
magnetism if temperature is lowered further. The diamagnetic volume
susceptibility for the magnetic field parallel to the {[}1 1 0{]}
direction (in this direction demagnetization correction is small \cite{Osborn_1945})
is greater than -0.5 indicating bulk superconductivity %
\footnote{$\chi_{v}$ is larger than the ideal value of -1. However, considering
the ferromagnetic contribution of the Eu$^{2+}$ (as found by the
scattering measurements), the effective volume susceptibility due
to the superconductivity (SC), ($\chi_{v}^{\textup{SC}}$ ) = $\chi_{\textup{observed}}-\chi_{\textup{Eu}}^{\textup{Ferro}}$,
might be very close to -1. The observed ferromagnetic contribution
for the non-superconducting ferromagnet EuFe$_{2}$P$_{2}$ \cite{Feng_2010}
(in ZFC data) is the same order of magnitude as the $\chi_{\textup{observed}}$
in the present case.%
}. Effective diamagnetic susceptibility close to -1 for the ZFC curve provides an upper limit of superconducting volume fraction of 100\%. Figures \ref{fig1} (c) \& (d) show hysteresis loops at \emph{T}
= 5 and 30 K for the two field directions. The observed hysteresis
curves look different than a type II nonmagnetic superconductor. However,
a jump in magnetization, which is typical for a type-II superconductor,
is clearly observed at 7 T magnetic field between the field increasing
and decreasing cycles. To understand the atypical hysteresis curve,
we assume a ferromagnetic contribution of the Eu$^{2+}$ moments at
an applied magnetic field \emph{H} (Tesla) by,
\begin{eqnarray}
m_{Eu} & = & (7.0/0.5)\times H\,\mu_{B},\,\textup{for}\,\,\left|H\right|\leq0.5\label{eq:meu}\\
 & = & 7.0\,\times H/\left|H\right|\mu_{B}\,\,,\,\textup{for\,\,}\left|H\right|\geq0.5\nonumber
\end{eqnarray}
since very little hysteresis was observed for the ferromagnetic end
member EuFe$_{2}$P$_{2}$ \cite{Feng_2010}. Lower insets to Fig.\,\ref{fig1}
(c) and (d) show magnetization after subtraction of the ferromagnetic
contribution from the Eu$^{2+}$ moments according to Eq.\,\ref{eq:meu}.
The hysteresis curves after subtraction look very similar to the other
Fe based superconductors \cite{Ruslan_08,Athena_08}. The jump at
7\,T magnetic field is consistent with Bean's critical state model
together with Lenz's law \cite{Bean_62,Kim_63,Fietz_64}. Reversal
of the direction of change of applied field as at 7 T does not remove
the specimen from the critical state but merely reverses locally the
direction of the critical current according to Lenz's law. Therefore,
magnetization measurements strongly hint towards a ferromagnetic superconductor
in an applied magnetic field. The heat capacity of the same single
crystal was measured using a Quantum Design physical property measurement
system (PPMS) and is shown in Fig.\,\ref{fig1}(e). Specific heat
data show a clear phase transition at \emph{T}$_{\textup{C}}$ = 19
K indicating the onset of the Eu$^{2+}$ magnetic order. A specific
heat jump at \emph{T}$_{\textup{SC}}$ is clearly visible and amounts
to $\Delta$C\,$\approx$ 350 mJ/mol.K which is slightly less but of
the same order of magnitude as that observed for the K-doped BaFe$_{2}$As$_{2}$
system \cite{Ni_08}. Due to the difficulties in determination of  $\Delta$C as well as "$\gamma$" as a result of large magnetic contribution at low temperatures, it will be hugely erroneous to estimate the value of $\Delta$C/($\gamma$\emph{T}$_{\textup{SC}}$) and make comparison with other non-magnetic iron based superconductors. Heat capacity measurement down to mK temperature range is needed to correctly estimate the value of $\gamma$. The entropy release associated with the magnetic
order of the Eu$^{2+}$ moments amounts to 17.1 J/mol.K which is equal
to 99\% of the expected theoretical value $R\ln(2S+1)$ for Eu$^{2+}$
moments with spin \emph{S}\,=\,7/2. Therefore, the specific heat
measurement indicates that substantial volume of the sample, if not 100\%, contributes to both the superconductivity and magnetic order of the Eu$^{2+}$ moments. Moreover, the
full moment of Eu$^{2+}$ is completely ordered at the single phase
transition temperature \emph{T}$_{\textup{C}}$ of 19 K.
\subsection{X-ray resonant magnetic scattering}
\begin{figure}
\includegraphics[width=0.42\textwidth]{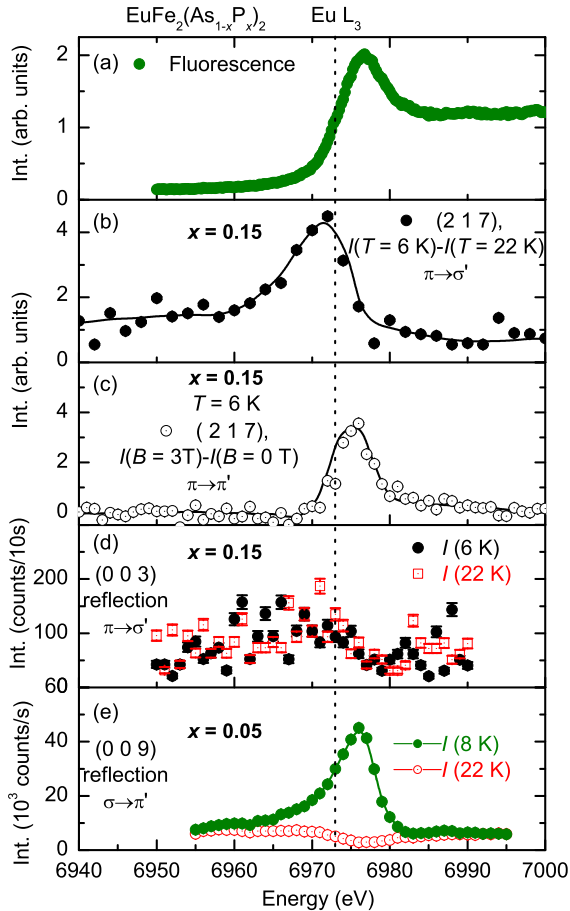}

\caption{\label{fig3}(a) Energy scan of the fluorescence yield. The dashed
line depicts the Eu L$_{3}$ absorption edge as determined from the
inflection point of the fluorescence yield. (b) and (c) Energy scans
for the (2 1 7) reflection after subtraction of the non-magnetic background
at high temperature for (b) and at zero magnetic field for (c). (d)
and (e) Energy scans through the antiferromagnetic ( 0 0 3) and (0
0 9) positions for the 15\% and 5\% samples, respectively. Lines serve
as guides to the eye.}
\end{figure}
To determine whether there is a structural phase transition, as observed
in the parent compound EuFe$_{2}$As$_{2}$, ($\xi$ $\xi$ 0)$_{T}$
scans were performed through the tetragonal (2 2 8)$_{T}$ Bragg reflection
as a function of temperature. The inset to Figure~\ref{fig2} (a)
shows a subset of ($\xi$~$\xi$~0)$_{T}$ scans through the (2~2~8)$_{T}$
reflection for the 15\% doped sample as the sample was cooled through
$T_{\textup{S}}$~=~49$\pm$1~K. The splitting of the (2\,2\,8)$_{T}$
Bragg reflection into orthorhombic ($O)$ (4\,0\,8)$_{O}$ and (0\,4\,8)$_{O}$
Bragg reflections below $T_{\textup{S}}$ is consistent with the structural
transition, from space group $I4/mmm$ to $Fmmm$, with a distortion
along the {[}1~1~0{]} direction. As the sample is cooled further,
the orthorhombic splitting ($\delta$) increases down to $T$~=~30$\pm$1~K
as can be seen from Fig.\,\ref{fig2}(a). Near \emph{T}$_{\textup{SC}}$,
$\delta$ shows a local minimum due to the competition between superconductivity
and ferromagnetism. Lowering the temperature below $T_{\textup{C}}$
results in a smooth decrease in $\delta$, reminiscent of that observed
in the superconducting Ba(Fe$_{1-x}$Co$_{x}$)$_{2}$As$_{2}$ samples
\cite{Nandi_PRL_10}. The non-superconducting 5\% doped sample undergoes
a similar structural phase transition at $T_{\textup{S}}$~=~165$\pm$1~K
but without any decrease of the orthorhombic distortion for lower
temperatures.

Below $T_{\textup{C}}$\,=\,20\,K, a magnetic signal was observed
when the x-ray energy was tuned through the Eu L$_{3}$ edge at reciprocal
lattice points identical to those of the charge reflections, indicating
the onset of the Eu$^{2+}$ magnetic order at the magnetic propagation
vector $\boldsymbol{\tau}$\,=\,(0\,0\,0). Figure~\ref{fig2}(b)
depicts the temperature evolution of the (2 1 7) reflection measured
at the Eu L$_{3}$ edge at resonance (\emph{E}\,=\,6.973\,keV).
A variation of the magnetic intensity with temperature was only observed
in the $\pi\rightarrow\sigma'$ scattering channel whereas the $\pi\rightarrow\pi'$
scattering channel shows no discernible temperature dependence. The
transition temperature is similar to that observed in the parent EuFe$_{2}$As$_{2}$
compound and consistent with the results presented in Fig.\,\ref{fig1}.
Figure~\ref{fig2}(c) shows temperature dependence of the same (2
1 7) reflection in an applied magnetic field of 0.5 T along the {[}1
$\overline{2}$ 0{]} direction in both scattering channels. It is
interesting to see that the temperature dependence appears in the
opposite scattering channel compared to the zero field and indicates
a possible flop of the magnetic moment in an applied magnetic field
which will be discussed later. The transition temperature is increased
from 19 K at zero field to 29 K at 0.5 T.

To confirm the resonant magnetic behavior of the peaks, we performed
energy scans at the Eu L$_{3}$ absorption edge as shown in Fig.~\ref{fig3}.
We note that for the (2\,1\,7) reflection charge and magnetic peak
coincide. An investigation of the magnetic signal which is five to
six orders of magnitude weaker than the Thomson charge scattering
requires significant reduction of the charge background. The charge
background can be reduced significantly for a reflection with scattering
angle close to 90$^{\circ}$ \cite{Nandi_PhysRevB.84.054419,kim:202501}.
Since the (2\,1\,7) reflection has a scattering angle of $\sim94.5{}^{\circ}$
at the Eu L$_{3}$ edge, the investigation of the magnetic signal
seems feasible for this reflection. Figure\,\ref{fig3}(b) shows
an energy scan through the (2\,1\,7) reflection after subtracting
the nonmagnetic background at \emph{T} = 22 K. A clear resonance enhancement
can be seen close to the Eu L$_{3}$ edge. A similar resonance enhancement
can be observed in the $\pi\rightarrow\pi'$ scattering channel in
an applied magnetic field of 3 T. In both energy scans, the resonance
peaks appear at and above the Eu L$_{3}$ absorption edge, indicating
the dipole nature of the transition. Figure \ref{fig3} (d) shows
energy scans through the antiferromagnetic (0 0 3) position, expected
for an A-type AFM structure, for the 15\% doped sample in the $\pi\rightarrow\sigma'$
scattering channel. For comparison, we also show the energy scan through
the (0 0 9) position in Fig.\,\ref{fig3}(e) for the 5\% doped sample
measured under similar conditions. A strong antiferromagnetic signal
was observed for the 5\% doped sample at the A-type AFM position which
is in contrast to the 15\% doped sample where no magnetic signal was
observed. Therefore, the proposed A-type AFM structure \cite{Zapf_11}
could not be confirmed for the superconducting 15\% P-doped sample.
This might be due to the small moment in the A-type AFM structure
together with the glassy freezing of the in-plane component as suggested
by Ref. \onlinecite{Zapf_13}.
\subsection{Magnetic structure in zero and applied magnetic fields}
\begin{figure*}
\includegraphics[width=0.92\textwidth]{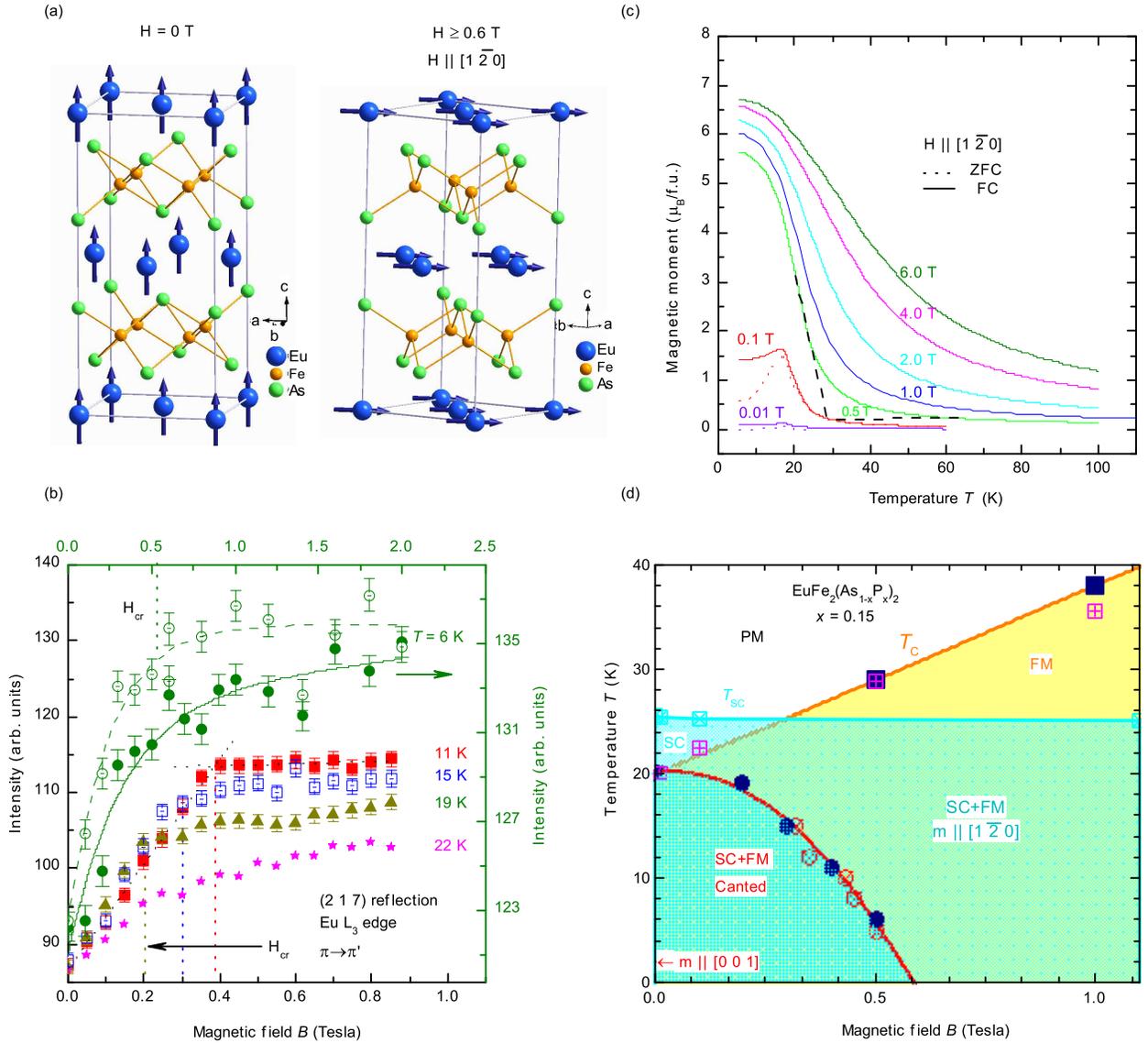}

\caption{\label{fig4}(a) Magnetic structures of the Eu$^{2+}$ moments in
zero and applied magnetic fields. Only the Eu$^{2+}$ magnetic moments
are shown. (b) Field dependence of the intensities of the (2 1 7)
reflection measured in the $\pi\rightarrow\pi'$ scattering geometry
after zero field cooling of the sample from 80 K. (c) Temperature
dependence of the bulk magnetization at different applied magnetic
fields along the {[}1 $\overline{2}$ 0{]} direction measured using
an MPMS. (d) Magnetic phase diagram for the 15\% doped sample constructed
using magnetization and scattering measurements. Filled symbols are
derived from the scattering measurement and the open symbols from
\emph{M}-\emph{T} (square) and \emph{M}-\emph{H} (circles) measurements
at different fields and temperatures, respectively. The transition temperatures, $T_{\textup{SC}}$ and $T_{\textup{C}}$, at zero field are consistent with the published results of Ref. \onlinecite{Jeevan_11}.}
\end{figure*}

\begin{table}[h]
\caption{Basis vectors for the space group $Fmmm$ with $\boldsymbol{\tau}$\,=\,(0\,0\,0).
The decomposition of the magnetic representation for the Eu site at
$(0\,0\,0)$ is $\Gamma_{Mag}=0\Gamma_{1}^{1}+0\Gamma_{2}^{1}+1\Gamma_{3}^{1}+0\Gamma_{4}^{1}+1\Gamma_{5}^{1}+0\Gamma_{6}^{1}+1\Gamma_{7}^{1}+0\Gamma_{8}^{1}$. }

\begin{ruledtabular} %
\begin{tabular}{ccccccc}
\multicolumn{1}{c}{IR } & Atom  & \multicolumn{3}{c}{BV components} & \multicolumn{2}{c}{Magnetic Intensity}\tabularnewline
\multicolumn{1}{c}{} &  & $m_{\|a}$  & $m_{\|b}$  & $m_{\|c}$  & \multicolumn{2}{c}{(2\,1\,7)}\tabularnewline
 &  &  &  &  & $\pi\rightarrow\sigma'$  & $\pi\rightarrow\pi'$ \tabularnewline
\hline
$\Gamma_{3}$  & 1  & 1  & 0  & 0  & Yes  & Yes \tabularnewline
$\Gamma_{5}$  & 1  & 0  & 1  & 0  & Yes & Yes \tabularnewline
$\Gamma_{7}$  & 1  & 0  & 0  & 1  & Yes  & No \tabularnewline
\end{tabular}\end{ruledtabular} \label{basis_vector_table_Eu}
\end{table}

We now turn to the determination of the magnetic moment configuration
for the Eu$^{2+}$ moments in the zero and applied magnetic fields.
For the crystallographic space group \emph{Fmmm} and $\boldsymbol{\tau}$\,=\,(0\,0\,0),
three independent magnetic representations (MRs) are possible \cite{Wills_00}.
Here we note that only ferromagnetic structures with magnetic moments
along the three crystallographic directions \textbf{a}\emph{, }\textbf{b}\emph{,
}\textbf{c} are allowed by symmetry. No antiferromagnetic structure
with $\mathbf{\boldsymbol{\tau}}$\,=\,(0\,0\,0) is possible in
this case for symmetry reasons. All the MRs along with the calculated
intensities for different polarization geometries are listed in Table\,
\ref{basis_vector_table_Eu}.

The resonant scattering of interest, at the Eu L$_{3}$ absorption
edge, is due to electric dipole transitions between the core 2\emph{p}
states and the 5\emph{d} conduction bands. The 5\emph{d} bands are
spin polarized through the exchange interaction with the magnetic
4\emph{f} electrons. The resonant magnetic scattering cross-section
for the dipole resonance can be written as \cite{Hill:sp0084}:
\begin{equation}
f_{nE1}^{XRMS}=[(\hat{\epsilon}^{'}\cdot\hat{\epsilon})F^{(0)}-i(\hat{\epsilon}^{'}\times\hat{\epsilon})\cdot\hat{z}_{n}F^{(1)}+(\hat{\epsilon}^{'}\cdot\hat{z}_{n})(\hat{\epsilon}\cdot\hat{z}_{n})F^{(2)}]\label{cross-section}
\end{equation}

where $\hat{z}_{n}$ is a unit vector in the direction of the magnetic
moment of the \emph{n}$^{th}$ ion. Here $\hat{\epsilon}$ and $\hat{\epsilon}^{'}$
are the incident and scattered polarization vectors, and $F^{(i)}$'s
are the terms containing dipole matrix elements. The first term of
Eq.~\ref{cross-section} contributes to the charge Bragg peak as
it does not contain any dependence on the magnetic moment. The other
two terms are sensitive to the magnetic moment. For a ferromagnetic
structure, in general all terms contribute to the scattering at every
Bragg reflection. However, for the Eu$^{2+}$ ions with spin only
magnetic moment, the spherical symmetry of the spin-polarized 5\emph{d}
band ensures that the \emph{F}$^{(2)}$ term is zero \cite{hamrick}.
For the $\pi\rightarrow\sigma'$ scattering geometry the scattering
amplitude from Eq.\,\ref{cross-section} can be written in a simplified
form as $f\propto\,\mathbf{k}_{i}\cdot\boldsymbol{\mu}$, \cite{Nandi_book}
where $\mathbf{k_{i}}$ and $\boldsymbol{\mu}$ are the wave vector
of the incoming photons and the magnetic moment, respectively. Clearly,
the magnetic signal is sensitive to the component of the ordered moment
in the scattering plane i.e. \emph{a}/\emph{b} and \emph{c} components.
For the $\pi\rightarrow\pi'$ scattering geometry the scattering amplitude
can be written as $f\propto\,(\mbox{\textbf{k}}_{i}\times\mathbf{k}_{f})\cdot\boldsymbol{\mu}$,
\cite{Nandi_book} where $\mathbf{k_{f}}$ is the wave vector of the
outgoing photons. Therefore, in the $\pi\rightarrow\pi'$ scattering
geometry, the magnetic signal is sensitive to the component of the
ordered moment perpendicular to the scattering plane \emph{i.e.} only
\emph{a}/\emph{b} components. Since, no magnetic signal was observed
in the $\pi\rightarrow\pi'$ scattering channel at zero field (see
Fig.\,\ref{fig2}(b)), we conclude that the magnetic moments are
aligned primarily along the \textbf{c} axis. For the applied magnetic
field the situation is reversed. The magnetic signal is observed only
in the $\pi\rightarrow\pi'$ scattering channel (see Fig.\,\ref{fig2}(c))
indicating the magnetic moments are in the \emph{a}-\emph{b} plane.
It is most likely that the magnetic moments are along the applied
filed direction \emph{i.e.} along the {[}1 $\bar{2}$ 0{]} direction.
The determined magnetic structures based on the polarization analysis
of the scattered signal is presented in Fig. \ref{fig4}(a).

Having determined the magnetic structures in zero and applied magnetic
fields, we have measured the field dependencies of the integrated
intensity of the magnetic (2 1 7) reflection for several temperatures
which are presented in Fig. \ref{fig4} (b). A clear hysteresis can
be seen from the increasing and decreasing field cycles at \emph{T}
= 6 K which is typical for a ferromagnet. The critical field, H$_{cr}$,
at which the field induced phase transition occurs, has been determined
from the intercept of the high and low field linear interpolation
as shown for the \emph{T} = 11 K measurement in Fig.\,\ref{fig4}
(b). The field dependence of the ferromagnetic ordering temperature
has been determined from the temperature dependence of the (2 1 7)
reflection in the $\pi\rightarrow\pi'$ scattering geometry as shown
in Fig. \ref{fig2}(c). Additionally, isothermal magnetization (\emph{M-H})
at different temperatures (not shown) and temperature dependencies
of magnetization (\emph{M-T}) at different magnetic fields (see Fig.
\ref{fig4}(c)) have been performed to verify the transition temperatures
and critical fields obtained from the scattering measurements. A combined
phase diagram has been constructed and is shown in Fig. \ref{fig4}(d).
It can be seen that superconductivity coexists with strong ferromagnetic
order of the Eu$^{2+}$ moments for a large region of the phase diagram.
For \emph{B} $\leq$ 0.5\,T, the superconducting transition precedes
the ferromagnetic transition, whereas the situation is reversed for
magnetic fields higher than 0.5 T.
\section{Discussion and Conclusion}
The most important result of the present study is the observation
of strong ferromagnetic order of the Eu$^{2+}$ moments coexisting
with bulk superconductivity. Magnetization, specific heat and temperature
dependence of the structural distortion indicates bulk nature of the
superconducting transition. In contrast to the previous studies, we
got no indication of the proposed A-type AFM structure or a spiral
magnetic order with propagation vector of the form (0 0 $\tau$) %
\footnote{Careful scans along {[}0 0 L{]} direction do not reveal any magnetic
peak.%
}. In the Fe-As based superconductors, it is believed that the superconducting
carriers are in the Fe-As layers. Therefore, to understand the phenomena
of coexistence, we have calculated the effective field due the Eu$^{2+}$
moments at the Fe-As layers using dipole approximation. To a first
approximation, the dipole field does not exceed 1\,T which is much
less than the superconducting upper critical field \emph{H}$_{\textup{C}2}$
(\textasciitilde{}\,40 T) \cite{Johnston} but higher than the lower
critical field \emph{H}$_{\textup{C}1}$ (\textasciitilde{}\,0.02-0.03
T) \cite{Athena_08}. Since the internal field is between \emph{H}$_{\textup{C}1}$
and \emph{H}$_{\textup{C}2}$, it is most likely that the EuFe$_{2}$(As$_{1-x}$P$_{x}$)$_{2}$
is in a spontaneous vortex state similar to which have been proposed
in Eu(Fe$_{0.75}$Ru$_{0.25}$)$_{2}$As$_{2}$ \cite{Jiao_11} and
UCoGe superconductors \cite{Deguchi_10}. At an applied magnetic field, it is most likely that the vortices in the zero-field state (along the \emph{c} axis) will gradually change along the applied field direction \emph{i.e.} in the \emph{a-b} plane.

In conclusion, the magnetic structure of the Eu moments in superconducting
EuFe$_{2}$(As$_{1-x}$P$_{x}$)$_{2}$ with $x=0.15$ has been determined
using element specific x-ray resonant magnetic scattering. Combining
magnetic, thermodynamic and scattering measurements we conclude that
the long range ferromagnetic order of the Eu$^{2+}$ moments aligned
primarily along the \textbf{c} axis coexists with the bulk superconductivity.
The proposed canted antiferromagnetic order or spiral order could
not be confirmed in the superconducting sample. Additional measurements
such as small angle neutron scattering is needed to confirm the existence
of a spontaneous vortex state.

\bibliographystyle{apsrev} \bibliographystyle{apsrev}
\begin{acknowledgments}
S. N. likes to acknowledge B. Schmitz and S. Das for technical assistance,
S. Zapf and M. Dressel for fruitful discussion. Work at Göttingen
was supported by the German Science Foundation through SPP 1458. \bibliographystyle{EuFeAs}
\bibliography{EuFeAs}
\end{acknowledgments}

\end{document}